\documentstyle[12pt,psfig,epsfig]{article}
\topmargin - 0.75cm
\begin{document}
\setlength{\parskip}{0.45cm}
\setlength{\baselineskip}{0.75cm}
%
%
%

\def\lapp{\raisebox{-.4ex}{\rlap{$\sim$}} \raisebox{.4ex}{$<$}}
\def\gapp{\raisebox{-.4ex}{\rlap{$\sim$}} \raisebox{.4ex}{$>$}}
\def\bra {\langle}
\def\ket {\rangle}
\def\lm {\lambda}
\def\g {\gamma}
\def\gm {\Gamma}
\def\dl {\delta}
\def\dg {\Delta \Gamma / \Gamma}
\def\r {\rightarrow}
\def\rb {\right}
\def\lb {\left}
\def\R {\longrightarrow}

\def\be {\begin{equation}}
\def\ee {\end{equation}}
\def\bea{\begin{eqnarray}}
\def\eea {\end{eqnarray}}
\def\n {\nonumber}
\def\bc{\begin{center}}
\def\ec {\end{center}}
\def\beq{\begin{equation}}
\def\eeq{\end{equation}}
\def\ba{\begin{array}}
\def\ea{\end{array}}
\def\bea{\begin{eqnarray}}
\def\eea{\end{eqnarray}}
\def\n{\nonumber}
\def\b{\bar}
\def\bq{\begin{quote}}
\def\eq{\end{quote}}

\def\bib{\bibitem}
\def\be{\begin{equation}}
\def\ee{\end{equation}}
\def\barr{\begin{array}}
\def\earr{\end{array}}
\def\dis{\displaystyle}
\def\eg{ {\em e.g.}}
\def\etc{ {\em etc.}}
\def\etal{ {\em et al.}}
\def\ie{ {\em i.e.}}
\def\viz{ {\em viz.}}
\def\Im{{\rm Im~}}
\def\Re{{\rm Re~}}
\def\lsim{\:\raisebox{-0.5ex}{$\stackrel{\textstyle<}{\sim}$}\:}
\def\gsim{\:\raisebox{-0.5ex}{$\stackrel{\textstyle>}{\sim}$}\:}
\def\rd{{\rm d}}
\def\mev{\: {\rm MeV} }
\def\gev{\: {\rm GeV} }
\def\tev{\: {\rm TeV} }
\def\pb{\: {\rm pb}}
\def\fb{\: {\rm fb}}
\def\ra{\rightarrow}
\def\mand{\qquad {\rm and} \qquad}
\def\ptsl{p_T \hspace{-1.1em}/\;}
\def\pslash{p \hspace{-0.6em}/\;}
\def\msl{m \hspace{-0.8em}/\;}

\def\gappeq{\mathrel{\rlap
  {\raise.5ex\hbox{$>$}}
  {\lower.5ex\hbox{$\sim$}}}}

\def\lappeq{\mathrel{\rlap{\raise.5ex\hbox{$<$}}
 {\lower.5ex\hbox{$\sim$}}}}
\def\ET {\not\!\!{E_T}}
\def\d {\delta}
\def\D {\Delta}
\def\f {\frac}
\def\l {\left}
\def\r {\right}
\def\G {\Gamma}
\def\o {\omega}
\def\e {\eta}
\def\lam {\lambda}
\begin{titlepage}
\setlength{\parskip}{0.25cm}
\setlength{\baselineskip}{0.25cm}
\renewcommand{\thefootnote}{\fnsymbol{footnote}}
\begin{flushright}
DO-TH 02/10\\
\vspace{0.2cm}
hep--ph/0209090\\
\vspace{0.2cm}
May 2002
\end{flushright}
\vspace{1.0cm}
\begin{center}
\LARGE
{\bf Does CPT violation affect the $B_d$}\\ 
\LARGE
{\bf meson life times and decay asymmetries?}
\vspace{1.0cm}

\large
Amitava Datta\footnote{Electronic address: 
adatta@juphys.ernet.in~.}\\
\vspace{0.25cm}

\normalsize
{\it Department of Physics, Jadavpur University,
Kolkata 700 032, India,}\\
\vspace{0.5cm}

\large
Emmanuel A.\ Paschos\footnote{Electronic address:
paschos@physik.uni-dortmund.de}\\
\vspace{0.25cm}

\normalsize
{\it Universit\"{a}t Dortmund, Institut f\"{u}r Physik,
D-44221 Dortmund, Germany,}\\
\vspace{0.5cm}

\large
and
\vspace{0.25cm}

\large
L.P.\ Singh\footnote{Electronic address:
lambodar@iopb.res.in}\\
\vspace{0.25cm}

\normalsize
{\it Department of Physics, Utkal University,
Bhubaneswar 751004, India}\\
\vspace{1.0cm}
\end{center}

\begin{abstract}
We study indirect CPT violating effects in $B_d$
meson decays and mixing, taking into account the recent 
constraints on the CPT violating parameters from the Belle collaboration. 
The life time difference 
of the $B_d$ meson mass eigenstates, expected to be negligible
in the standard model and many of its CPT conserving extensions,
could be sizeable ($\sim$ a few percent of the total width) due to
breakdown of this fundamental symmetry. The time evolution of the  direct
CP violating asymmetries in one amplitude dominated processes (inclusive
semileptonic $B_d$ decays, in particular) turn out to be particularly
sensitive to this effect. 
\end{abstract}
\end{titlepage}



The suggestion for two distinct lifetimes for the $B_d$ or $B_s$ 
meson mass eigenstates originated 
in parton model calculations \cite{ref1}, which, at 
that time, were limited by numerous uncertainties
of hadronic ($f_B$, the bag parameter, top quark mass, 
...) and weak parameters (CKM matrix elements). Many of 
these, however, cancel in the ratio
\begin{equation}
\left(\frac{\Delta m}{\Delta\Gamma}\right)_d = 
 \frac{8}{9\pi}
  \left( \frac{\eta_t}{\eta}\right)\,
   \left(\frac{m_t}{m_b}\right)^2 f(x_t)
\end{equation}
where $\Delta m_d(\Delta\Gamma_d)$ is the mass (width) 
difference of the $B_d$ meson mass eigenstates,
$\eta_t,\,\eta$ are calculable perturbative  QCD 
corrections, $x_t = \frac{m_t}{m_w}$ and
\begin{equation}
f(x) = \frac{3}{2} \frac{x^2}{(1-x)^3}\, \ln x-
 \left( \frac{1}{4}+\frac{9}{4}\frac{1}{(1-x)}\, -
  \frac{3}{2}\, \frac{1}{(1-x)^2}\right)\, .
\end{equation}

Following the discovery of mixing in the $B_d$ system
\cite{ref2}, $\Delta m_d$ was measured and $m_t$ was
the only major source of uncertainty in the ratio. 
Using the then lower bound on $m_t$ it was shown 
\cite{ref3,ref4} that $\Delta\Gamma_d$ is indeed very 
small, while $\Delta\Gamma_s$, the width difference 
of the $B_s$ meson mass eigenstates could be rather 
large as is indicated by the scaling law \cite{ref4}
\begin{equation}
\left(\frac{\Delta\Gamma}{\Gamma}\right)_s =
  \left( \frac{X_{Bs}}{X_{Bd}}\right)\cdot
    \left|\frac{V_{ts}}{V_{td}}\right|^2\cdot
\left(\frac{\Delta\Gamma}{\Gamma}\right)_d
\end{equation}
where $V_{ij}$'s are the elements of the CKM matrix and
\begin{equation}
X_{B_q}=\langle B_q|\left[ \bar{q}\gamma^{\mu}
  (1-\gamma_5)b\right]^2|\bar{B}_q\rangle\, .
\end{equation}

In the meanwhile, many advances have taken place with 
the discovery of the top quark and the determination 
of its mass \cite{ref5} and more precise values for 
CKM matrix elements. Combining the new values with the
above scaling laws, the width difference among the 
$B_d$ states is $(\Delta\Gamma/\Gamma)_d \approx 0.0012$, 
which is unobservable, but for $B_s$ eigenstates 
$(\Delta\Gamma/\Gamma)_s\approx 0.045$.
More recent calculations using heavy quark effective 
theory and improved QCD corrections \cite{ref6,ref7} 
suggest that calculations  based on the absorptive
parts of the box diagram improved by QCD corrections 
give reasonable estimates for both $B_d$ and $B_s$ 
systems.

Nevertheless the possibility that there are loopholes 
in the above calculations cannot be totally excluded. 
For example, $\Delta \Gamma_q$ (q = d or s) is determined by only 
those  channels which are accessible to both $B_q$ and 
$\bar{B}_q$ decays. Its computation in the parton model 
may not be as reliable as the calculation of 
$\Gamma_q$, the total width which depends on  fully 
inclusive decays and quark- hadron duality is valid.

In addition to the expected phenomena, one should, 
therefore, be prepared for unexpected effects and the 
final verdict on this subject should wait for
experimental determination  of $\Delta\Gamma$ from the 
B--factories, B--TeV or LHC--B.  Many different 
suggestions for measuring $\Delta\Gamma_s$ have been put 
forward \cite{ref3,ref4,ref8}.  It is believed that 
$(\dg)_d \sim  0.1$ can be measured at B--factories 
\cite{ref9} while $(\dg)_d \sim 0.001$ \cite{ref10} 
might be accessible at the LHC.

In this article we wish  to emphasize that 
apart from dynamical surprises in the decay mechanism, 
a possible breakdown of the CPT symmetry  contributes 
to $\dg$. The currently available constraints on CPT 
violating parameters \cite{opal,belle} certainly allow  this 
possibility. If this happens its effect will be more
visible and detectable in the $B_d$ system which, in 
the electroweak theory, is expected to have negligible 
$(\dg)_d$. In other words the scenario with $(\dg)_d$ large 
not only due to hitherto unknown dynamics but also due 
to a breakdown of CPT is quite an open possibility. In 
the case of  $(\dg)_s$ CPT violation may act in tandem 
with the already known electroweak dynamics to produce
an even larger effect.\\

There are several motivations for drawing out a strategy 
to test CPT symmetry. From  the experimental point of 
view all symmetries of nature must be scrutinized as
accurately as possible, irrespective of the prevailing 
theoretical prejudices. It may be recalled that before 
the discovery of CP violation, there was very little 
theoretical argument in its favour.

There are purely theoretical motivations as well. First 
of all the CPT theorem is valid for local, renormalizable 
field theories with well defined asymptotic states. It 
is quite possible that the theory we are dealing with 
is an effective theory and involving small nonlocal/ 
nonrenormalizable interactions. Further the concept of 
asymptotic states is not unambiguous in the presence 
of confined quarks and gluons. It has been suggested 
that physics at the string scale may indeed induce
nonlocal interactions in the effective low energy 
theory leading to CPT violation \cite{ref11}. Moreover, 
modification of quantum mechanics due to gravity may 
also lead to a breakdown of CPT \cite{ref12}.

One of the major goals of the B--factories running at
KEK or SLAC is to reveal CP violation in the B system.
The discrete symmetry CPT has not yet been adequately 
tested  for the B meson system, although there 
are many interesting suggestions to test it \cite{ref13,ref14}.
 In all 
such works, however, the correlation between 
$\D \gm$  and CPT violation was either ignored or not 
adequately emphasized. It will be shown below that  
$\D \gm$ can in general be numerically significant 
even if  CPT  violation is not too large.

We consider the time development of neutral mesons
$M^0$ (which can be $K^0$ or $D^0$ or $B_d^0$ or
$B_s^0$) and their antiparticles $\bar{M}^0$.
The time development is determined by the effective Hamiltonian
 $H_{ij} = M_{ij}-\frac{i}{2}\Gamma_{ij}$ 
with $M_{ij}$ and 
$\Gamma_{ij}$ being the dispersive and absorptive 
parts of the Hamiltonian, respectively \cite{ref15}.  
  CPT invariance relates the diagonal elements
\begin{equation}
M_{11} = M_{22}\quad\quad {\rm and} \quad\quad
  \Gamma_{11} = \Gamma_{22}\, .
\end{equation}
A measure of CPT violation is, therefore,  given by the parameter
\begin{equation}
\delta = 
 \frac{H_{22}-H_{11}}{\sqrt{H_{12}H_{21}}}
\end{equation}
which is phase convention independent.
In order to keep the discussion simple we shall
study the consequences of indirect CPT violation
only.  
Since indirect CPT violation is a cumulative effect
involving summations over many amplitudes, it is
likely that its magnitude would be much larger than
that of direct violation in a single decay amplitude.
It is further assumed that CPT violation does not
affect the off--diagonal elements of $H_{ij}$. These
assumptions can be justified in specific string models
\cite{ref11}, where terms involving both flavour and
CPT violations receive negligible corrections due
to string scale physics.  A further consequence of
this assumption is that the usual SM inequality 
$M_{12}\gg \Gamma_{12}$ holds even in the presence
of CPT violation. 

The eigenfunctions of the Hamiltonian are defined as
\begin{equation}
|M_1\rangle = p_1|M^0\rangle + q_1|\bar{M}^0\rangle
 \quad\quad {\rm and} \quad\quad
|M_2\rangle = p_2|M^0\rangle -q_2|\bar{M}^0\rangle
\end{equation}
with the normalization $|p_1|^2+|q_1|^2 = |p_2|^2+
 |q_2|^2=1$.  We summarize the consequences of the 
symmetries.  We define
\begin{eqnarray}
\eta_1 = \frac{q_1}{p_1} =
 \left[\left(1+\frac{\delta^2}{4}\right)^{1/2} +
   \frac{\delta}{2}\right]\,
     \left[\frac{H_{21}}{H_{12}}\right]^{1/2}\\
\eta_2 = \frac{q_2}{p_2} =
  \left[\left(1+\frac{\delta^2}{4}\right)^{1/2} -
   \frac{\delta}{2}\right]\,
      \left[\frac{H_{21}}{H_{12}}\right]^{1/2}
\end{eqnarray}
and note that CPT violation is contained in the first
factor, while indirect CP violation is in the second
factor with the square root.  In many expressions we
need the ratio  
$\omega=\eta_1/\eta_2 =\frac{q_1 p_2}{p_1 q_2}$
which is only a CPT violating quantity.  CPT
conservation requires\\ Im $\omega=0$,  Re $\omega=1$ and
$\eta_1=\eta_2$.

The time development of the states is determined by
the eigenvalues
\begin{eqnarray}
\lambda_1 & = & H_{11}+\sqrt{H_{12}H_{21}}
 \left[ \left( 1+\frac{\delta^2}{4} \right)^{1/2} 
  + \delta/2 \right] \quad {\rm and}\nonumber\\
\lambda_2 & = & H_{22}-\sqrt{H_{12}H_{21}}
 \left[ \left( 1+\frac{\delta^2}{4}\right)^{1/2}
  + \delta/2 \right] 
\end{eqnarray}
which can be parametrized as $\lambda_{1,2} =
m_{1,~2}-\frac{i}{2}~\Gamma_{1,~2}$.  The quantities 
which occur in the asymmetries are 
$\lambda_1-\lambda_2 = 
\Delta m -\frac{i}{2}\Delta\Gamma$ and $\Gamma =
\frac{1}{2}(\Gamma_1+\Gamma_2)$. To leading order
in $(\Gamma_{12}/M_{12})$ 
they are expressed in terms of the CPT parameter
\begin{displaymath}
y= \left(1+\frac{\delta^2}{4}\right)^{1/2}
\end{displaymath}
as follows:
\begin{eqnarray}
\Delta m = m_1 - m_2 = 2|M_{12}|(\Re\, y+\frac{1}{2}\,
    \Re \frac{\Gamma_{12}}{M_{12}}\Im\,y)\nonumber\\
\Delta\Gamma=\Gamma_1-\Gamma_2=2|M_{12}|
 (\Re\, \frac{\Gamma_{12}}{M_{12}}\, \Re\, y -2\Im\, y)\, .
\end{eqnarray}
In the CPT conserving limit $y=1$ and the contribution
to $\Delta m$ is large, overwhelming CPT violating
corrections.  The CPT conserving contribution to 
$\Delta\Gamma$, on the other hand, is suppressed
by  $\Re\, \frac{\Gamma_{12}}
{M_{12}}$. The purely CPT violating term dominating
$\Delta \Gamma$ remains, therefore, an open possibility.
In order to get a feeling for the magnitude of 
$\Delta\Gamma/\Gamma$, we use the small $\delta$
approximation and obtain $|\Delta\Gamma/\Gamma|=0.5
\times (\Delta m/\Gamma)\times (\Re\delta\times 
\Im\delta)$. 

Most of the  measurements of
$\Delta m/\Gamma$ have  been carried out
by assuming CPT conservation.  If CPT is violated
its magnitude could be somewhat different (see, e.g.,
Kobayashi and Sanda in \cite{ref13}). Recently the Belle collaboration 
has determined $\Delta m$ with and without assuming CPT symmetry 
\cite{belle}.
 The two results
, $\D m$ = 0.463 $\pm$0.016 and 0.461$\pm$0.008 $\pm$ 0.016 $ps^{-1}$
respectively, do not differ appreciably from each other or from
the average $\D m$ given by the particle data group(PDG).
We shall, therefore,  use throughout
the paper $\Delta m/\Gamma=0.73$, which is perfectly consistent with 
the PDG value. The relevant limits on CPT violating parameters
from Belle are $|m_{B^0} - m_{\bar{B^0}}|/m_{B^0} < $1.6 $\times 10^{-14}$
and $|\gm_{B^0} - \gm_{\bar{B^0}}|/\gm_{B^0} <$ 0.161, which implies $|\Re
\dl
| < $ 0.54 and $|\Im \dl| <$ 0.23.  
A choice like $\Re\delta \times  \Im\delta\sim 0.1$, consistent with the
above bounds,
 would then yield $\Delta\Gamma/\Gamma$ of the order of a few \%,
 larger than the SM estimate by an order of
magnitude.  Moreover $\Delta\Gamma/\Gamma$ will be 
well within the measurable limits of LHC B,  should 
$\delta$ happen to be much smaller.

The Belle limits are derived under the assumption that $\dg$ is 
negligible.
 We emphasize that for a refined analysis of CPT violation
$\Delta m/\Gamma$ and $\Delta\Gamma/\Gamma$ along 
with $\delta$ should be fitted directly from the data. Such a combined fit 
may open up the possibility that $\dl$ could be somewhat larger than
the above bounds. In our numerical analysis values consistent with the
bounds as well as somewhat larger values will be considered.

The time development of the states involves, now, 
the time factors
\begin{eqnarray}
f_-(t) & = & e^{-i\lambda_1t}-e^{-i\lambda_2t}\\
f_+(t) & = & e^{-i\lambda_1t}+
   \omega e^{-i\lambda_2t}\quad{\rm and}\\
\bar{f}_+(t) & = & \omega e^{-i\lambda_1t} +
    e^{-i\lambda_2t}\, .
\end{eqnarray}
The new feature is the presence of the factor 
$\omega$ in the second and third of these equations.

The decays of an original $|B^0\rangle$ or 
$|\bar{B}^0\rangle$ state to a flavor eigenstate
$|f\rangle$ vary with time and are given by
\bea P_f(t) = \l|< f | B^0(t)>\r|^2
            =\l|f_+(t)\r|^2 N \l|<f|B^0>\r|^2 \eea
\bea \bar P_{\bar f}(t) = \l|<\bar f |\bar B^0(t)>\r|^2
            =\l|\bar f_+(t)\r|^2 N 
\l|<\bar f|\bar B^0>\r|^2 \eea
\bea  P_{\bar f}(t) = \l|<\bar f | B^0(t)>\r|^2
            = \l|\eta_1\r|^2  \l| f_-(t)\r|^2 N \l|<\bar 
f|\bar
B^0>\r|^2
\eea
\bea \bar P_{f}(t) = \l|<f |\bar B^0(t)>\r|^2
            =\l| f_-(t)\r|^2 N \l|\omega\r|^2 
\l|<f|B^0>\r|^2/\l|\eta_1\r|^2
\eea
where $N^{-1}$ = $|1 + \omega|^2$ and  
the matrix elements on the right--hand side
$ g = \langle f|B^0\rangle, \, \bar{g}=\langle \bar {f}|\bar{B}^0
\rangle,\, \ldots$, are computed at $t=0$ and have
no time dependence. From these expressions it is
evident that the five unknowns, $\Gamma,\, \Delta m,
\,  \Delta\Gamma$ and $\Re\delta$ and $\Im\delta$ (or
equivalently $\Re\omega$ and $\Im\omega$), must be
determined from the data.  We emphasize that 
$\Delta\Gamma$ must be treated as a free parameter,
since in addition to the CPT violating contributions
it may also receive contributions from new dynamics.
In addition taking linear combinations of these
decays, we can produce exponential decays accompanied
by oscillatory terms which help in separating the
various contributions. It may be recalled that the
time dependent techniques
for extracting these probabilities and the associated electroweak 
parameters from data are now being used extensively.

Different schemes  for testing CPT violation
suggested in the literature \cite{ref13,ref14} often involve 
 observables specifically constructed for this purpose. 
  Here we  wish to point out that some of the observables
 involving the above probabilities,
which are now being routinely measured at BABAR and BELLE are also
sufficiently sensitive to CPT violation and have the potential
of either  revealing the breakdown of this fundamental symmetry
or  improving the limit on the CPT violating parameter.  
One such observable is the direct CP violating asymmetry 
in $B_d$ and  $\bar{B_d}$ decays to flavor specific channels 
$f$ and $\bar{f}$, respectively \cite{ref16},  
but with $f$ different from $\bar{f}$. 
The following ratio is at the center of current interest 
\bea
{\mbox{ a}}^{\mbox{dir}}_{CP} (t)
       = \f { \l|<f|B^0(t)>\r|^2 - \l|<\bar f|\bar B^0(t)>\r|^2}
            { \l|<f|B^0(t)>\r|^2 + \l|<\bar f|\bar B^0(t)>\r|^2}  \n\\ 
= \f { \l|f_+(t)\r|^2 \l|g\r|^2 - \l|\b f_+(t)\r|^2 
  \l|\b g \r|^2 }
  { \l|f_+(t)\r|^2 \l|g\r|^2 + \l|\b f_+(t)\r|^2 
\l|\b g \r|^2}
\eea

In the SM or in any of its CPT conserving extensions,
 $\b f_+(t) = f_+(t)$ and  the asymmetry is time independent in
general.  The time independence holds  even if 
  $\D \gm$  happens to be large  due to new dynamics
or   direct CPT violation and/ or new physics  influence the hadronic 
matrix elements.  Time evolution of this asymmetry is,
therefore, a sure signal of indirect CPT violation. Flavour specific
B decays involving a single lepton or a kaon in the final state are
possible candidates for this measurement.

This consequence is even more dramatic for decays
dominated by a single amplitude in the SM, in which case 
$|g|=|\bar{g}|$ and ${\mbox{ a}}^{\mbox{dir}}_{CP} (t)$ vanishes
at all times.
Purely tree level decays arising from the subprocess
$b\to u_i\bar{u}_j d_k$ ($i \ne j$), penguin induced processes
$b\to d_i\bar{d}_i d_k$, dominated by a single Penguin
operator or inclusive semileptonic decays 
$b\to X l^+ \nu $ (l = e or $\mu$ and X is any hadronic final  state) are
examples of such decays. The last process is particularly promising.
A single  amplitude  strongly dominates the decay  not only in the
SM but also in many extensions of it. 
The large branching ratio ($\sim 20\%$ for l = e and $\mu$ ) and
reasonably large efficiency of detecting  leptons is
sufficient to ensure the measurement of this  asymmetry at B - factories,
 provided it is of the order of a few percent. 

For this class of decays the  matrix elements along with their theoretical
uncertainties cancel out in the ratio.  Consequently
in presence of indirect CPT violation the time dependent  asymmetry is the
same for all one--amplitude dominated processes and the statistics may be 
improved by including several channels.  If a difference in the 
time dependence of various modes is observed, the
assumption of  one amplitude dominance  will be
questionable and new physics beyond the standard model leading to
$|g| \ne |\bar{g}|$, in addition to indirect CPT violation, 
may be revealed.

In Figure 1 we present the asymmetry for a one 
amplitude dominated process as a function of time for Im $\dl$ = 0.1
and Re $\dl$ = 0.1 (solid curve, here $\dg $ = 0.004)
, 0.5 (dotted curve, $\dg$ = 0.02) and 1.0 (dashed curve, $\dg$=0.04). 
Both the time evolution
and the nonvanishing of the asymmetry are clearly demonstrated.

The correlation between  ${\mbox{ a}}^{\mbox{dir}}_{CP}$ and $\D \Gamma$
 calls for a more detailed analysis. 
As has been noted $\D \Gamma$ is significantly different from the SM
prediction  only if $\Im \dl~ \times ~ \Re \dl \ne 0 $.
The numerator and the denominator of the asymmetry are  determined to be
\bea
D(t)& = &  P_f(t) -  {\bar P}_{\bar f}(t)\nonumber\\
  &   = &\lb [ \lb (\lb |\omega \rb |^2 - 1\rb) \lb ( e^{-\Gamma_2t}
- e^{-\Gamma_1 t}\rb )-4~{\rm Im}~ \omega~  e^{-\Gamma t} \sin\Delta m 
t  
\rb ] N
\eea
and 
\bea
S(t)& = &  P_f(t) +  {\bar P}_{\bar f}(t)\nonumber\\
  &   = &\lb [ \lb (\lb |\omega \rb |^2 + 1\rb) \lb ( e^{-\Gamma_2t}
+ e^{-\Gamma_1 t}\rb )+4~{\rm Re}~\omega  e^{-\Gamma t} \cos\Delta m 
t  
\rb ] N
\eea
A non-vanishing asymmetry can arise in various ways\\
            i) Im $\o \ne$0, which requires Im$\delta \ne$0.0,\\
           ii) Re $\o \ne$ 1.0 and $\dg$ as small as in the SM, \\
or from a combination of the two possibilities. It is trivial to
express $\omega$ in
terms of $\delta$ and confirm that both
D(t) and S(t)
are modified from the SM prediction through $\delta$. When both numerator
and denominator of the asymmetry are measured accurately, one can
determine separately real and imaginary parts of delta. This may
indicate, albeit indirectly, that $\D \Gamma$ is unexpectedly large.

In order to have  an idea of how large the effects
can be,
in Figure 2 we plot $D(t)$
as a function of time for the values $\Re~ \delta=\Im~
\delta=0.1$ (the solid curve).  
D(t)  vanishes for $\Im\delta=0$
and has a relatively weak  dependence on $\Re\delta$, as 
illustrated by also plotting on the same figure the cases 
with $\Re\delta=0.5$ (the dotted curve) and 1.0 (the dashed
curve). A similar study of S(t) is presented in Figure 3.
This quantity is fairly  insensitive to Im $\dl$.  

In order to estimate roughly the number of tagged B-mesons needed to
establish a  non-zero D(t)
, we assume that at t=0  there is a sample of $N_0$ tagged
$B_d^0$ and $\bar{B}_d^0$. Let of number of semileptonic $B_d^0$
($\bar{B}_d^0$ ) decays in the time interval
t=$ (1.0 \pm 0.1) \times \tau_B$ be n(t) ($\bar{n}(t)$)
(we  assume the lepton detection efficiency to be $\sim$ 1).
 By requiring
\begin{equation}
\frac{|n(t) - \bar{n}(t)|}{\sqrt{n(t)} + \sqrt{\bar{n}(t)}} \geq 3.0 , 
\nonumber
\end{equation}
\noindent
we obtain for Re $\delta$ = 0.1 and Im $\delta$ = 0.1, $N_0
 \approx $ 2.0
$\times$ 10$^6$, a number which is realizable  at B - factories
after several years of run  
and certainly at the LHC.  Including other flavour specific
channels like $B_d^0  \rightarrow K^+ + X$, which has a larger branching
ratio ($\approx$ 70\%), a measurable asymmetry may be obtained with a
smaller $N_0$.

In the presence of indirect CPT violation,
the  time integrated asymmetry is obtained by integrating the
numerator D(t) and the denominator S(t). This leads to
\bea a_{\rm CP}^{\rm dir}  =
\lb(  \lb(  \lb |\omega\rb |^2-1\rb)\frac{\Delta\Gamma}{\Gamma}
-\frac{4~{\rm Im}~\omega~ x}{\lb(1+x^2\rb)}\rb)\mbox{\Huge /}
\lb(2\lb(\lb |\omega\rb |^2+1\rb)+\frac{4~{\rm 
Re}~\omega}{\lb(1+x^2\rb)}
\rb)
\eea
with $x~ =~ \Delta m~ /~ \Gamma$.
In the standard model and for processes dominated by one amplitude
the integrated asymmetry vanishes. In extensions of the SM in which
the decays are no longer dominated by one amplitude the integrated
asymmetry may be nonzero \cite{ref17}. 
Thus a nonzero integrated asymmetry points either
to new physics (coming from additional amplitudes) or to indirect CPT
violation. In Figure 4  we  present the variation of this observable
with 
Im $\dl$ for Re $\dl$ = 0.1 (solid line) and 0.75 (dashed line).\\

Experimental studies of CPT violating phenomena can
be combined with experiments that search for a
$\Delta\Gamma/\Gamma$.  For example, one can consider
untagged B mesons decaying to a specific flavour
\cite{ref3,ref4}. The observable 
\bc

$S_1(t) =  P_f(t)+ \bar P_{f}(t)$
\ec 
 which in the absence of CPT 
violation have a time dependence governed by two
exponentials.  If now CPT violation is also included, 
then 
an oscillation is superimposed on the exponentials.
The original articles \cite{ref3,ref4} considered
$B_s$ decays but the same properties hold for $B_d$
meson decaying semileptonically or to specific flavour
final states.
 
Looking at flavour non-specific channels there are 
 results for $a^{\rm dir}_{CP}$ (also denoted by $C_{\pi \pi}$)
from BABAR \cite{ref18}
and BELLE \cite{ref19} for the channel $B\to \pi^+\pi^-$. In the SM
using naive factorization this asymmetry turns out  to be small 
\cite{ref20}. It is
interesting to note that although the Babar result is fairly 
consistent with the SM prediction, the BELLE  result indicates a much
larger asymmetry. It should, however, be noted that there are
many theoretical uncertainties. Neither the magnitude of the penguin
pollution nor the magnitude of the strong phase difference 
between the interfering amplitudes can be
computed in a full proof way. Direct CP violation in flavour specific,
charmless decays have also been measured \cite{ref21}. Here the data
 is not yet 
very precise and the theoretical uncertainties are also large.
In view of these uncertainties it is difficult to draw any conclusion
regarding new physics effects. 
This underlines the importance of inclusive semileptonic decays
which are theoretically clean and the branching ratios are much  larger 
than any of the above exclusive modes.

$B_d$ decays to CP eigenstates have been observed
and established CP--violation via time dependent 
measurements \cite{ref22,ref23}. The golden example
is $B^0\to\psi K_s$ where the time dependent asymmetry is 
proportional to $\sin 2 \beta$, where $\beta$ (also denoted by  $\phi_1$) 
is an 
angle of the unitarity triangle. The current averaged value 
of this parameter is  $\sin 2 \beta$ =0.78 $\pm$ 0.08.             . 
It is straight--forward to obtain the  asymmetry in the presence
of indirect CPT violation. An attempt to fit the data as in the SM 
would lead to an
effective sin $2 \beta$ which is time dependent.
We have checked that with  Re $\delta$ = 0.1 and Im 
$\delta$= 0.1, this  effective sin $2 \beta$ 
 varies between 0.74 and 0.84. We therefore
conclude that if sin 2$\beta$ is determined with an accuracy
of 5 \% or better, some  hint of indirect CPT violation may
be obtained. 
However, this observation cannot establish CPT
violation unambiguously. Since CPT conserving new physics
may change the phase of the $B_d - \bar{B_d}$  mixing amplitude
and/or the decay amplitudes and lead to similar effects.\\

Many other observables specifically constructed  for the 
measurement of CPT violation 
\cite{ref13,ref14} have  been suggested  in the literature. 
It will be interesting to compare 
the sensitivities of these observables to the CPT parameter 
$\dl$  with that of  the observables
 considered in this paper, which are already being measured
in the context of CP violation. 

In summary, we wish to emphasize again that an  unexpectedly large life
time difference of $B_d$ mesons, which is predicted to be negligible 
in 
the SM and many of its CPT conserving extensions, may reveal 
indirect CPT violation. Time dependence of 
the direct CP violating asymmetry for flavour specific decays, 
which is time independent and vanishes for decays dominated by only
one amplitude  may establish CPT violation as well as a 
large life time difference. The theoretically clean inclusive
semileptonic decays having relatively large branching ratios 
might be particularly suitable  in this context.  
\noindent{\large\bf{Acknowledgements}}\\
    We wish to thank the Bundesministerium fur Bildung and and Forschung
for financial support under contract No. 05HT1PEA9. One of us (EAP)
thanks Mr. W. Horn for useful discussions. AD thanks the Department of
Science and Technology, Government of India for financial support under
project no SP/S2/k01/97 and Abhijit Samanta for help in computation.

\def\lapp{\raisebox{-.4ex}{\rlap{$\sim$}} \raisebox{.4ex}{$<$}}
\def\gapp{\raisebox{-.4ex}{\rlap{$\sim$}} \raisebox{.4ex}{$>$}}
\def\bra {\langle}
\def\ket {\rangle}
\def\lm {\lambda}
\def\g {\gamma}
\def\gm {\Gamma}
\def\dl {\delta}
\def\dg {\Delta \Gamma / \Gamma}
\def\ra {\rightarrow}
\def\R {\longrightarrow}

\def\im{{\rm Im}}
\def\re{{\rm Re}} 
\def\lsim{\:\raisebox{-0.5ex}{$\stackrel{\textstyle<}{\sim}$}\:}
\def\gsim{\:\raisebox{-0.5ex}{$\stackrel{\textstyle>}{\sim}$}\:}
\def\rd{{\rm d}}

\def\mand{\qquad {\rm and} \qquad}
\def\d {\delta}
\def\D {\Delta}
\def\f {\frac}
\def\l {\left}
\def\r {\right}
\def\o {\omega}
\def\e {\eta}
\def\lam {\lambda}

\newpage

\begin{figure}[htb]
\centerline{
\psfig{file=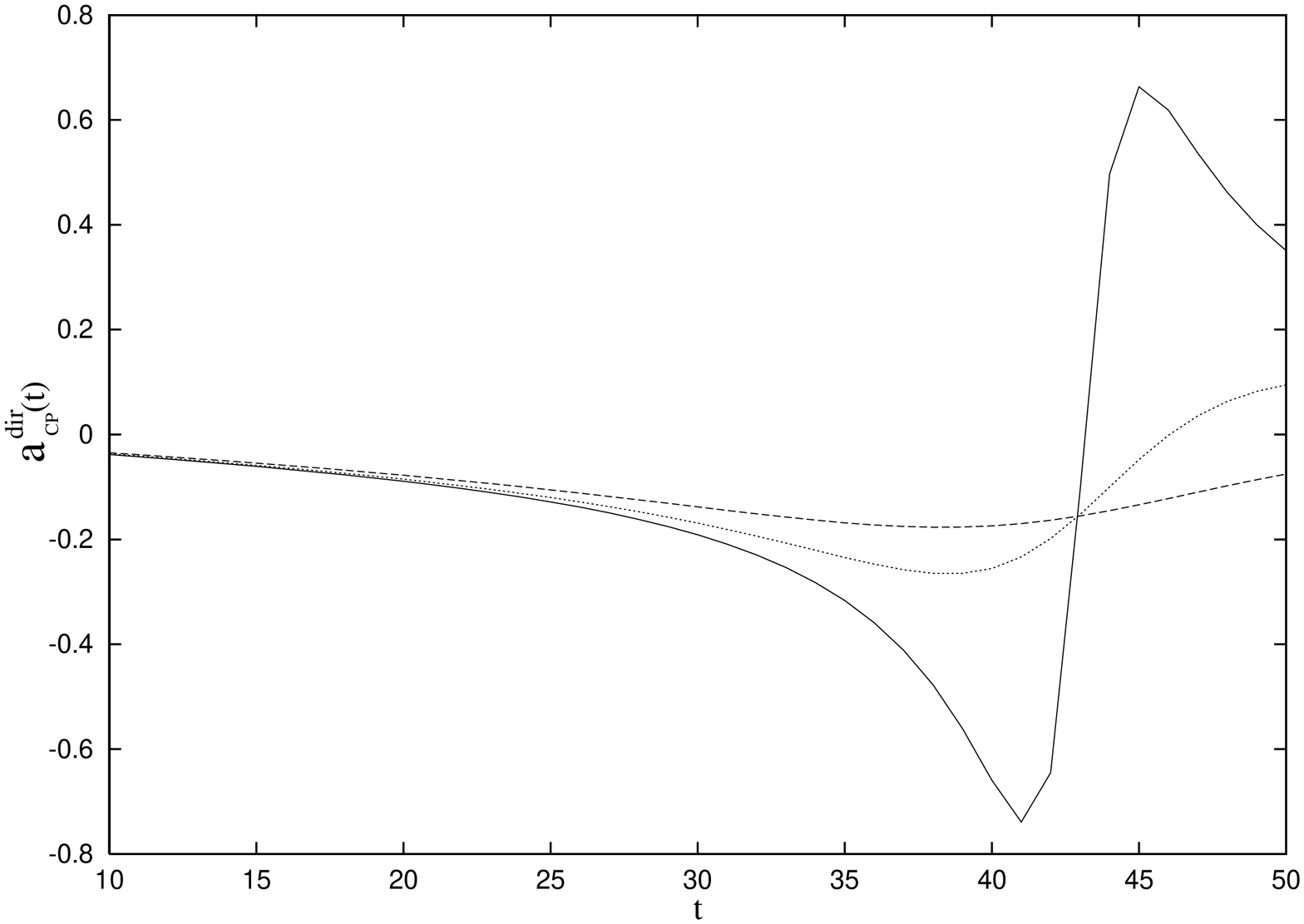,width=7cm,height=11cm,angle=270}
}
\caption{\sl{
The time evolution of the time dependent asymmetry ($a^{dir}_{CP}
$(t)) for $\im \dl$ = 0.1 for any one amplitude dominated process
(in all figures t is in units of 
$\tau_B/10$, where $\tau_B$ is  the average $ B^0$ life time ). 
See text for the details.            }}
\centerline{
\psfig{file=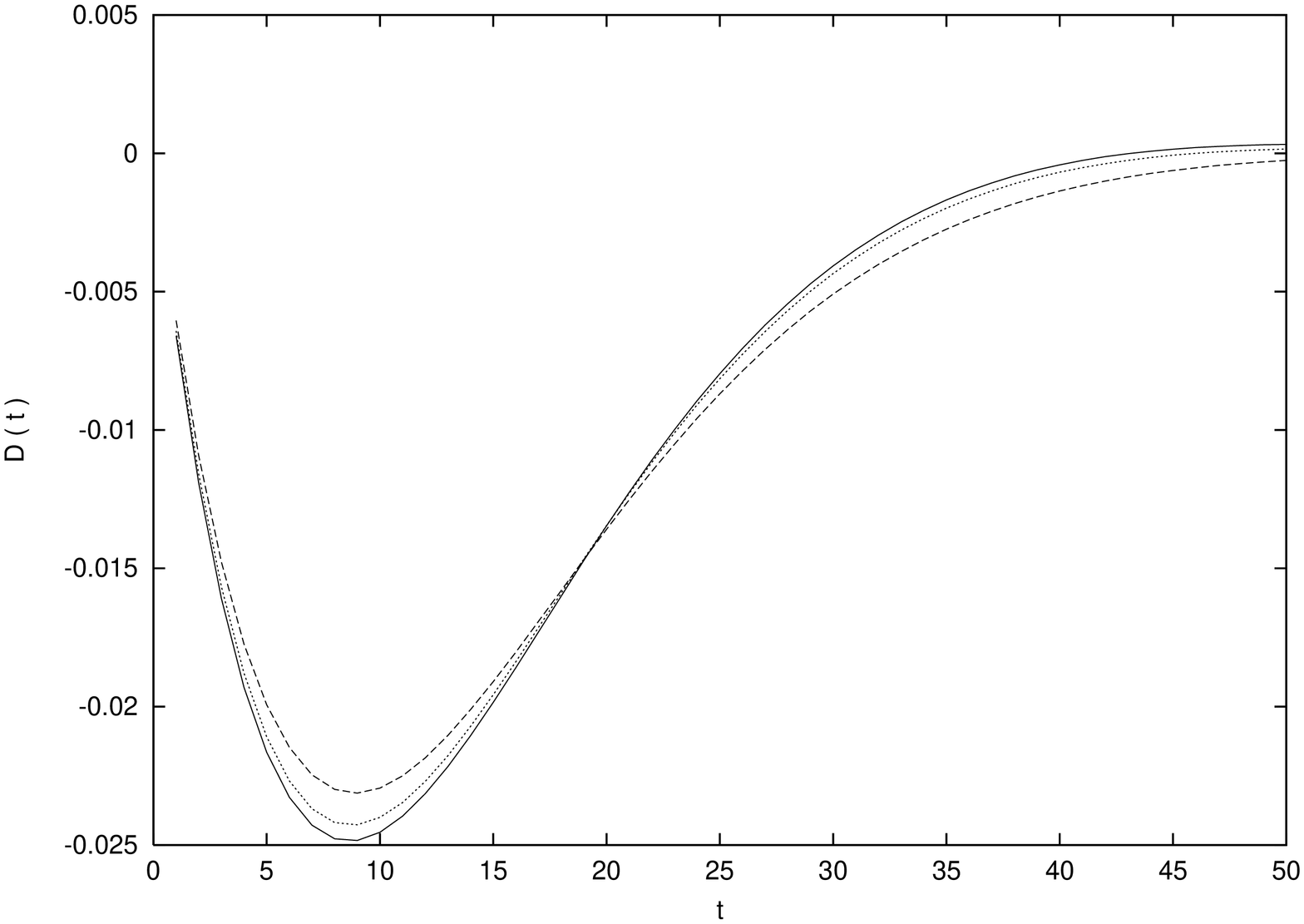,width=7cm,height=11cm,angle=270}
}
\caption{\sl{
The variation  of D(t) = Prob($B \ra f$) -  Prob($\bar B \ra \bar f$) 
as function of time for $\im \dl$ = 0.1 for any one amplitude 
dominated process. See text for the details. 
            }}
\end{figure}

\begin{figure}[htb]
\centerline{
\psfig{file=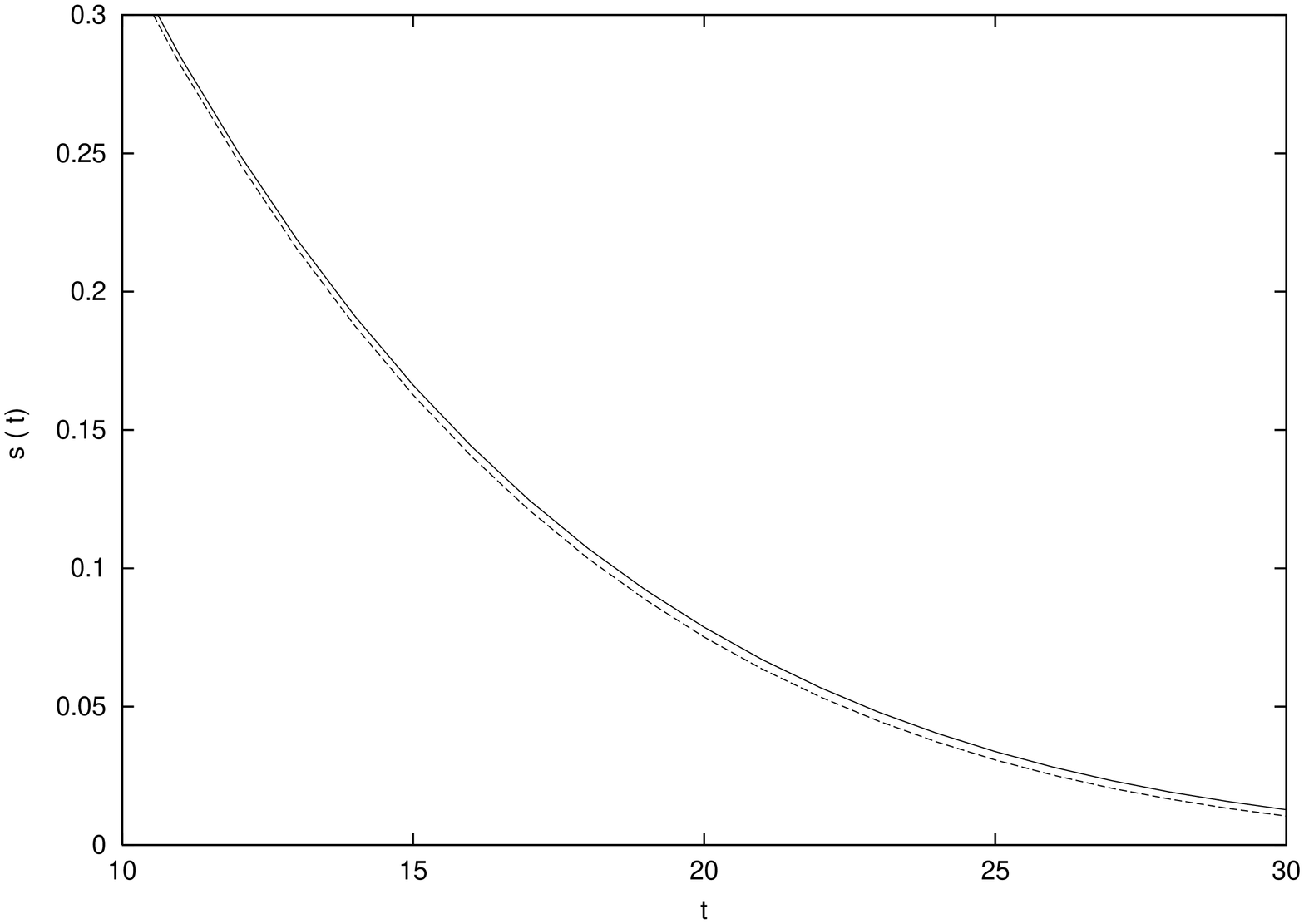,width=7cm,height=11cm,angle=270}
}
\caption{\sl{
The variation  of S(t)= Prob($B \ra f$) +  Prob($\bar B \ra \bar f$) 
as function of time. This quantity is sensitive to $\re \; \dl$
only. For comparison we have plotted for $\re \dl$=0.0 (SM)
(the dotted curve)  and
$\re \dl$ = 0.5 (the solid curve). 
            }}

\centerline{
\psfig{file=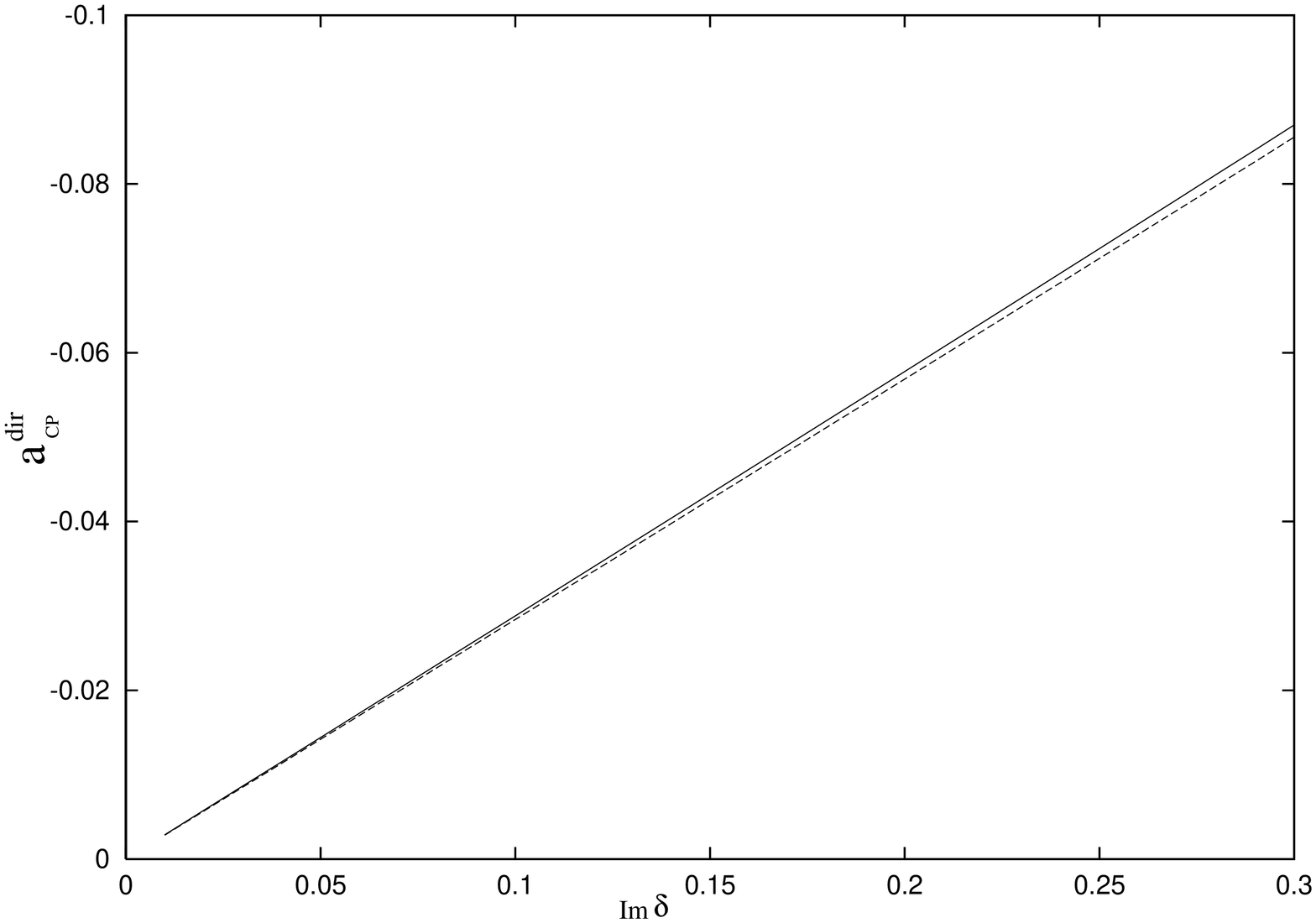,height=11cm,width=7cm,angle=270}
}
\caption{\sl{
The variation  of the  time integrated asymmetry $a^{dir}_{cp}$ for a one
amplitude dominated process vs  $\im \dl$. See text for further details.
            }}
\end{figure}  

\begin{thebibliography}{54}
\bibitem{ref1} J.S.~Hagelin, {\it Nucl.~Phys.}
        {\bf B193} (1981) 123;\\
        V.A.~Khoze, M.A.~Shifman, N.G.~Uraltsev and
        M.B.~Voloshin, {\it Sov.\ J.\ Nucl.\ Phys.\ Fiz.}
        {\bf 46} (1987) 112
\bibitem{ref2} H.~Albrecht et al.\ (ARGUS collaboration),
        {\it Phys.~Lett.} {\bf B192} (1987) 245
\bibitem{ref3} A.~Datta, E.A.~Paschos and U.~T\"urke,
        {\it Phys.~Lett.} {\bf B196} (1987) 382
\bibitem{ref4} A.~Datta, E.A.~Paschos and Y.L.~Wu,
        {\it Nucl.~Phys.} {\bf B311} (1988) 35
\bibitem{ref5} S.~Abachi et al.\ (CDF collaboration),
        {\it Phys.~Rev.~Lett.} {\bf 74} (1995) 2632
\bibitem{ref6} R.~Aleksan, A.~Le Yaouanc, L.~Oliver and
        J.C.~Raynal, {\it Phys.\ Lett.} {\bf B316} (1993)
        567
\bibitem{ref7} M.~Beneke et al.\ {\it Phys.~Lett.}
        {\bf B459} (1999) 631;\\
        A.S.~Dighe et al., {\it Nucl.~Phys.} {\bf B624} (2002)
377
\bibitem{ref8} I.~Dunietz, {\it Phys.\ Rev.}
        {\bf D52} (1995) 3048;\\
        I.~Dunietz, R.~Fleischer and U.~Nierste,
        {\it Phys.\ Rev.}
        {\bf D63} (2001) 114015;\\
        A.S.~Dighe et al.\ in ref 7
\bibitem{ref9} R.~Aleksan, private communications
\bibitem{ref10} A.S.~Dighe et al., in ref. 7
\bibitem{opal} K. Ackerstaff{\it et al}
 (OPAL collaboration) , {\it Z. Phys.} {\bf C76} (1997) 401.
\bibitem{belle} C. Leonidopoulos (Belle collaboration), hep-ex/0107001 and 
Ph.D thesis, Princeton University, 2000 (see http:$//$
belle.kek.jp)
\bibitem{ref11} V.A.~Kostelecky and R.~Potting,
        {\it Phys.~Lett.} {\bf B381} (1996) 89
\bibitem{ref12} S.W.~ Hawking, {\it Phys. Rev.}
        {\bf D14} (1976) 2460;\\
        J.~Ellis et al., {\it Phys.\ Rev.}
        {\bf D53} (1996) 3846
\bibitem{ref13} M.~Kobayashi and A.~Sanda, {\it Phys.~Rev.~Lett.}
        {\bf 69} (1992) 3139;\\
        Z.Z.~Xing, {\it Phys.~Rev.} {\bf D50}
        (1994) 2957;\\
        D. Colladay and V.A.~Kostelecky, {\it Phys.~Lett.}
{\bf B344} (1995) 359;\\
        V.A.~Kostelecky and R.~Potting, {\it Phys.~Rev.}
        {\bf D51} (1995) 3923;\\
        V.A.~Kostelecky and R.~Van Kooten {\it Phys.~Rev.}
        {\bf D54} (1996) 5585;\\
        M.C.~Banuls and J.~Bernabeu, {\it Nucl. Phys.} {\bf B590}
        (2000) 19;\\
        P .~Colangello and G.~Corcella, {\it Eur. Phys. J.}
        {\bf C1} (1998) 515;\\
A. Mohapatra et al., {\it Phys.~Rev.}
        {\bf D58} (1998) 036003. 
\bibitem{ref14} K.C.~Chou, W.F.~Palmer, E.A.~Paschos
        and Y.L.~Wu, {\it Eur.~Phys.~J.} {\bf C16}
        (2000) 279
\bibitem{ref15} See for instance, E.A.~Paschos and
        U.~T\"urke, {\it Phys.~Rep.} {\bf 178} (1989)
        145
\bibitem{ref16} For a review see, e.g., I.I. Bigi and A. Sanda, 
CP violation, Ed. C. Jarlskog (World Scientific, 1989); 
I.I. Bigi and A. Sanda, ``CP violation'' 
(Cambridge Univrsity Press, 1999). 
\bibitem{ref17}R-parity violating models are interesting in
this context. See, e.g.,  G.~Bhattacharya and A.~Datta,
        {\it Phys.~Rev.~Lett.} {\bf 83} (1999) 2300.
\bibitem{ref18}  Talk by A. Farbin (BABAR collaboration), XXXVIIth
Rencontres de Moriond, Les Arcs, France, 9 - 16 March, 2002,
http:$//$ moriond.in2p3.fr/EW/2002/.
\bibitem{ref19} K. Abe {\it et al} (BELLE collaboration), hep-ex/0204002.
\bibitem{ref20} A. Ali, G. Krammer and C-D. Lu, {\it Phys.~Rev.}
        {\bf D59} (1998) 014005.
\bibitem{ref21} B. Aubert {\it et al} (BABAR collaboration),
 {\it Phys.~Rev.}
        {\bf D65} (2002) 051101.
\bibitem{ref22}  Talk by T. Karim (BELLE collaboration), XXXVIIth
Rencontres de Moriond, Les Arcs, France, 9 - 16 March, 2002,
http:$//$moriond.in2p3.fr/EW/2002/.
\bibitem{ref23} B. Aubert {\it et al} (BABAR collaboration), hep-ex/0203007. 
\end{thebibliography}
\end{document}